# Multicolor Graphene Nanoribbon/Semiconductor Nanowire Heterojunction Light-Emitting Diodes

*Yu Ye,*[†] *Lin Gan,*[‡] *Lun Dai,*[*,†] *Hu Meng,*[†] *Feng Wei,*[†] *Yu Dai,*[†] *Zujin Shi,*[‡] *Bin Yu,*[†] *Xuefeng Guo,*[‡] *and Guogang Qin*[*,†]

† State Key Lab for Mesoscopic Physics and School of Physics, Peking University, Beijing 100871, China

‡ College of Chemistry and Molecular Engineering, Peking University, Beijing 100871, China

## Abstract

We report novel graphene nanoribbon (GNR)/semiconductor nanowire (SNW) heterojunction light-emitting diodes (LEDs) for the first time. The GNR and SNW have a face-to-face contact structure, which has the merit of bigger active region. ZnO, CdS, and CdSe NWs were employed in our case. At forward biases, the GNR/SNW heterjunction LEDs could emit light with wavelengths varying from ultraviolet (380 nm) to green (513 nm) to red (705 nm), which were determined by the band-gaps of the involved SNWs. The mechanism of light emitting for the GNR/SNW heterojunction LED was discussed. Our approach can easily be extended to other semiconductor nano-materials. Moreover, our achievement opens the door to next-generation display technologies, including portable, "see-through", and conformable products.

---

Corresponding author. E-mail: lundai@pku.edu.cn (L. D.); qingg@pku.edu.cn (G. Q.).

## Introduction

Graphene, a single atomic layer of carbon arranged in a honeycomb lattice, has attracted a lot of research interest since its discovery in 2004.[1-5] Its many important physical properties, such as high mobility and conductivity,[6] high optical transparency,[7] mechanical flexibility[8] and robustness,[9] and environmental stability,[9] have made graphene a promising material in diverse device applications. For example, it has been used to substitute indium titanium oxide (ITO) as the flexible transparent conductive electrode for organic photovoltaics[10] and light-emitting diodes (LEDs).[11] Besides, graphene is considered as a promising alternative for silicon for next-generation nanoelectronics, especially in producing low-cost, high-efficiency, lightweight, transparent, and flexible devices.[5, 12-16] So far, a lot of work has been done regarding the above two aspects. However, few work[17] aimed at developing the graphene-based nano-optoelectronic devices, which can be used in next-generation integratred nano-optoelectronic system, has been reported.

Semiconductor single crystalline nanowires (NWs) can be grown on basically any substrates, and can be constructed into devices with the bottom-up method.[18] Moreover, they are mechanically flexible and robust, compatible with low-temperature device process desired for plastic substrates.[19] So far, they have been used as building blocks in diverse devices, including Si-based optoelectronic devices,[20] and flexible transparent electronic/photonic devices.[19, 21]

In this letter, we report, for the first time, the graphene nanoribbon (GNR)/semiconductor NW (SNW) heterojunction LEDs. *n*-type ZnO, CdS, CdSe NWs were used in this work. A face-to-face contact structure was employed in the GNR/SNW heterojunction, which has the merit of larger active area, where the radiative recombination occurs.[22] The emitting light wavelength was determined by the involved SNW. Our achievement can easily be extended to other semiconductor nano-materials, and has a promising application on transparent flexible devices. Moreover, our approach pioneers a new realm for graphene-based high efficient electroluminescence (EL) devices, and even electrically driven lasers, which have potential application in developing the graphene-based micro/nano-optoelectronic integration in the future.

**Experimental Section**

Both the *n*-type SNWs[23-25] and the graphene[26] used in this work were synthesized via the CVD method. Before device fabrication, the graphenes were transferred by the stamp method with the help of PMMA (poly methyl methacrylate)[27] to Si/300 nm $SiO_2$ substrates for Raman and electrical property characterizations, and transferred to carbon-coated grids for high-resolution transmission electron microscopy (HRTEM) characterization (Tecnai F30). Their electrical properties were measured by a Hall-effect measurement system (Accent HL5500).

The fabrication processes of a GNR/SNW heterojunction LED are as follows (Figure 1): First, the as-synthesized large-scale graphene was transferred by the stamp method with the help of PMMA[27] to a Si/$SiO_2$ substrate. Second, the SNWs (ZnO, CdS, or CdSe here) dispersed in ethanol were dropped on the substrate (Figure 1a). Third, a photoresist pad was patterned to cover one end of a SNW by UV lithography and development processes (Figure 1b). The photoresist together with the uncovered SNW was then used as the masks for the following graphene etching process by an inductively coupled plasma (ICP) etching technique (Figure 1c). Later, the rest photoresist was removed by acetone, leaving a graphene pad connecting with a GNR underneath the SNW. Ohmic contact electrodes to the SNW (In/Au (10 /100 nm)) and graphene pad (Au (60 nm)) were defined by UV lithography followed by thermal evaporation and lift-off processes (Figure 1d). It is worth noting that because an undercut was form during the oxygen plasma etching process,[22] the In/Au electrode on the SNW will not contact with the GNR to cause a short circuit. The electrical properties of the GNRs fabricated with the method mentioned above were investigated by measuring their transistor characteristics. A GNR transistor was fabricated by using a $SiO_2$ NW as the mask. Here, the insulating $SiO_2$ NW needs not to be removed before the electrical measurement, and thus avoids additional chemical or physical processes induced influence to the measured results.

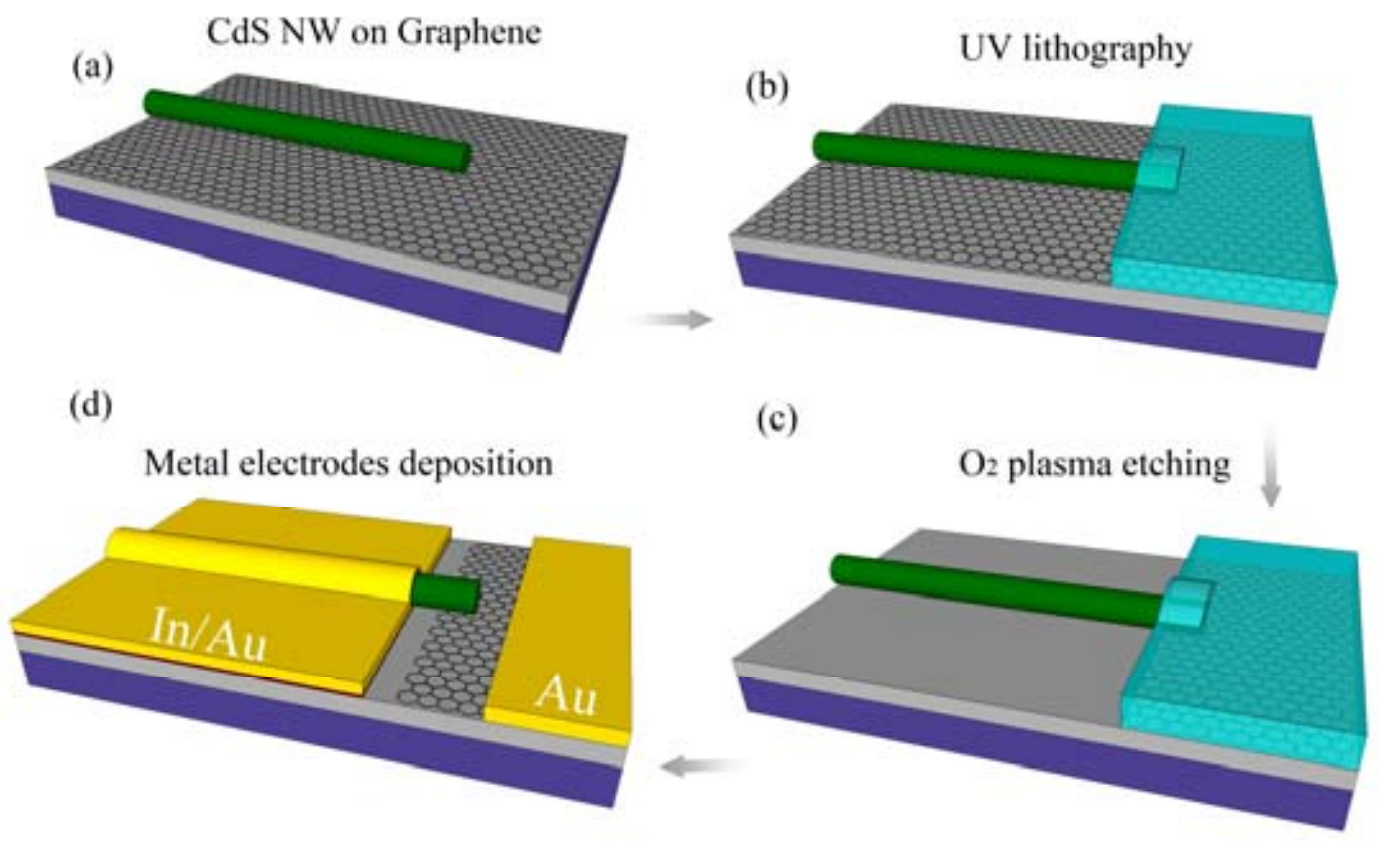

Figure 1. Schematic illustration of the fabrication processes of a GNR/SNW heterojunction LED. (a). The as-synthesized large-scale graphene was transferred to a Si/$SiO_2$ substrate. After that, SNW suspension was dropped on the graphene. (b). A photoresist pad was patterned to cover one end of a SNW by UV lithography and development processes. (c). Oxygen plasma etching was used to remove the exposed graphene. (d). After removing the photoresist, In/Au and Au ohmic contact electrodes to SNW and graphene were defined, respectively.

Room-temperature electrical transport properties of the GNR transistors and GNR/SNW heterojunctions were characterized with a semiconductor characterization system (Keithley 4200). Raman and EL measurements were done with a microzone confocal Raman spectroscopy (HORIBA Jobin Yvon, LabRam HR 800) equipped with a color charge-coupled device (CCD).

## Results and discussion

The HRTEM and Raman characterization results for the as-synthesized graphenes (see Supporting Information) demonstrate that the graphenes are high quality and with monolayer. The typical sheet resistance, hole concentration, and hole mobility of the graphene obtained by Hall-effect measurement are about 345 Ω/□, $1.84\times10^{14}$ $cm^{-2}$, and 98.6 $cm^2/V\cdot s$, respectively. It is worth noting that the low sheet resistance and high hole concentration of the as-synthesized graphene guarantee a small series resistance of the GNR in the heterojunction LED.

Figure 2a shows a field-emission scanning electron microscope (FESEM) image of an as-fabricated GNR/CdS NW heterojunction LED. We can see that the diameter of the CdS NW is about 300 nm. During the electrical measurement, the In/Au ohmic contact electrode of the CdS NW was grounded. The *I-V* curve (Figure 2b) of the LED shows an excellent rectification characteristic. An on/off current ratio of $\sim3.4\times10^7$ can be obtained when the voltage changes from +1.5 to –1.5 V. The turn-on voltage is around 1.1 V. In view of the high hole concentration ($1.84\times10^{14}$ $cm^{-2}$) and near-zero band-gap of the GNR,[28] the heterojunction structure of the GNR/CdS NW can be considered approximately as a metal-semiconductor contact of Schottky model.[29, 30] For Schottky barrier diodes made on high-mobility semiconductors such as ZnO, CdS and CdSe etc., the current $I$ is determined by the thermionic emission of electrons and can be described by the equation $I = I_0[\exp(eV/nkT)-1]$,[31] where $I_0$ is the reverse saturation current, $e$ is the electronic charge, $V$ is the applied bias, $n$ is the diode ideality factor, $k$ is the Bolzmann's constant, and $T$ is the absolute temperature. By fitting the measured *I-V* curve with the above equation, we can obtain $n$=1.58. Note that, the GNR/ZnO NW and GNR/CdSe NW heterojunctions show similar rectification characteristics as above, with the turn-on voltages to be about 0.7 and 1.2 V, respectively.

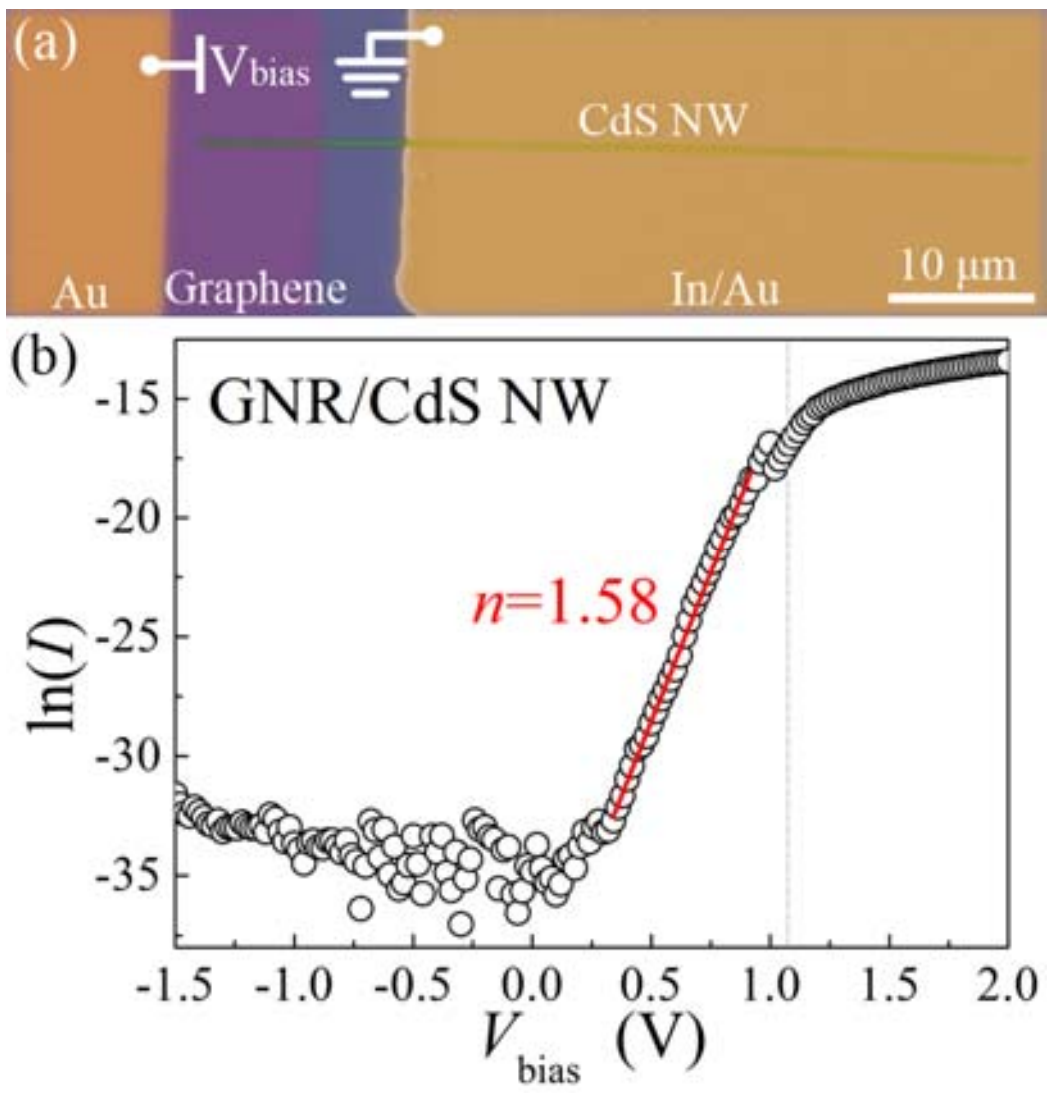

Figure 2. (a) FESEM image of an as-fabricated GNR/CdS NW heterojunction LED. (b) Room-temperature *I-V* characteristic of the LED in panel (a) on a semi-log scale. The red straight line shows the fitting result of the *I-V* curve by the equation $\ln(I) = eV/nkT + \ln(I_0)$.

Figures 3a-c show the EL images (Olympus BX51M) of the GNR/SNW (ZnO, CdS, CdSe, respectively) heterojunction LEDs at a forward bias of 5 V. Except for the ZnO NW case (where the emitting light is invisible ultraviolet light) in Figure 3a, strong emitting light spots can be seen clearly with naked eyes at the exposed ends of the NWs. For the CdS NW case (Figure 3b), we can see another glaring light spot on the NW. This may be due to the scattering from defect or adhered particle on the CdS NW.[32] As the optical image of the graphene on the Si substrate with 600 nm $SiO_2$ layer is unclear, we use the dashed lines to demarcate the graphenes in these figures. Figures 3d-f show the room-temperature normalized EL spectra at various forward biases for the GNR/SNW heterojunction LEDs, where the SNWs are ZnO, CdS, and CdSe NWs, respectively. The EL light collecting points were at the corresponding exposed ends of the NWs, indicated by the white arrows in the figures. The peak wavelengths of the EL spectra coincide with the band-edge emission of the involved SNWs (ZnO, CdS, CdSe, respectively.). This indicates the radiative recombinations of electrons and holes occur in the SNWs. For all the LEDs, EL intensities increase with the forward biases. Significantly, light emission can be detected at a forward bias as low as 2.5 V for the CdS NW case. We have studied more than 40 GNR/SNW heterojunction LEDs. Similar results are obtained.

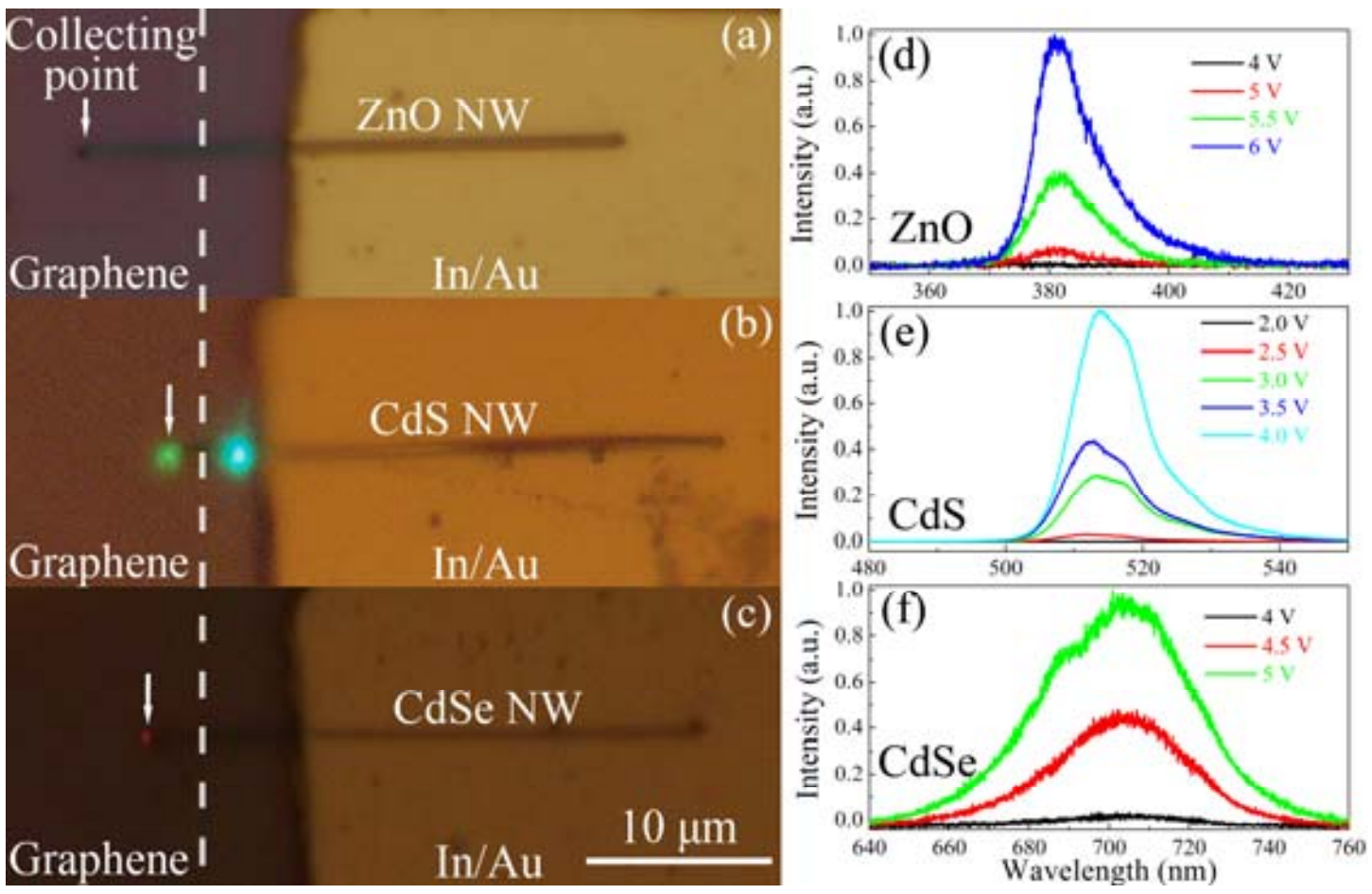

Figure 3. (a)–(c). The optical images of the GNR/SNW (ZnO, CdS, CdSe, respectively) heterojunction LEDs at a forward bias of 5 V. Dashed lines were used to demarcate the graphenes from the substrates. White arrows: the light collecting points during the EL measurements. (d)-(f). The room-temperature EL spectra for GNR/SNW (ZnO, CdS, CdSe, respectively) heterojunction LEDs at various forward biases.

Various Si-based SNW heterojunctions have been reported before.[20, 23] Lieber's group even reported the electrically driven laser from the Si/CdS NW heterojunction for the first time.[33] Compared with Si, graphene has higher conductivity and mobility,[6] which might make the graphene-based SNW heterojunciton superior to its Si-based rival in realizing high efficient LED or even laser device.

We can qualitatively understand the mechanism of the light emitting for the GNR/SNW heterojunction LEDs by studying the energy band diagrams. Figure 4a shows the thermal equilibrium energy band diagram of a graphene/*n*-type semiconductor structure, where the work function of graphene is $\Phi$, and the electron affinity of the semiconductor is $\chi$. $E_g$, $E_F$ correspond to the band-gap and Fermi level of the semiconductor. Due to the difference of their work functions, the energy band of semiconductor will bend upward at the graphene/semiconductor interface, and the Fermi levels at the two sides are brought into coincidence after contacting. Under a forward bias (*i.e.* a positive bias on graphene), the built-in potential is lowered. Therefore, more electrons will flow from *n*-type semiconductor to graphene, and simultaneously, more holes will flow from graphene to *n*-type semiconductor. Herein, as the SNW is the direct band-gap semiconductor, which has higher electron-hole radiative recombination rate, the injected holes and electrons will mainly recombine in the SNW region. Therefore, the corresponding EL spectrum will be determined by the band-edge emission of the SNW. It is worth noting that, in our face-to-face contact LED, the active region, where the radiative recombination occurs, is bigger, compared to the crossed nanowire structure reported previously.[20, 23, 34, 35] This merit would benefit the realization of high efficient EL, or even electrically driven laser of GNR/SNW heterojunction in future. For this purpose, further optimization of the SNW and graphene materials, the device structure, as well as the fabrication processes is ongoing.

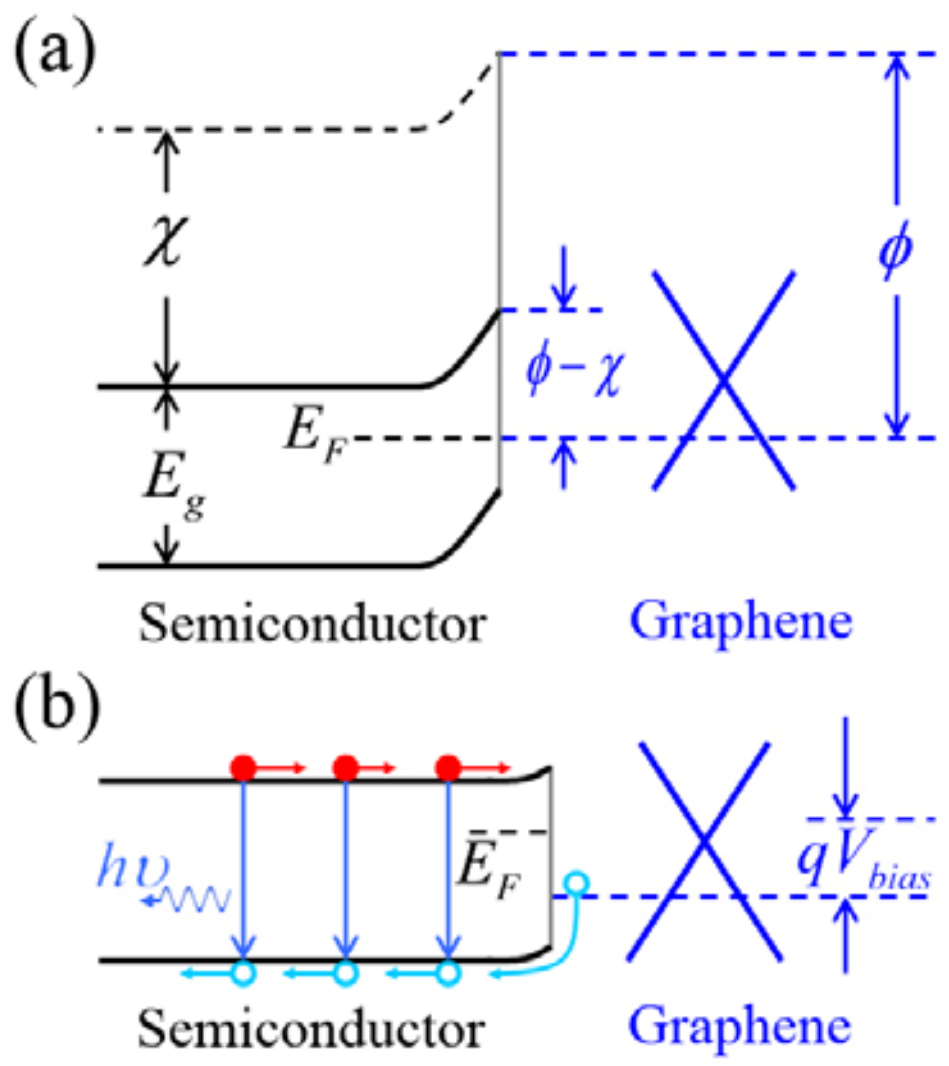

Figure 4. Schematic illustration of the energy band diagram of a graphene/semiconductor heterojunction. (a) The thermal equilibrium energy band diagram. (b) The energy band diagram of the heterojunction under a forward bias. $\Phi$: the work function of graphene; $\chi$: the electron affinity of the semiconductor.

## Conclusion

We have fabricated and studied GNR/SNW heterojunction LEDs for the first time. The GNR and SNW have a face-to-face contact structure, which has the merit of bigger active region. ZnO, CdS, and CdSe NWs were employed in our case. The emitting light wavelengths of the as-fabricated LEDs are determined by the band-gaps of the involved SNWs in the GNR/SNW heterojunctions, which can vary from ultraviolet (380 nm) to green (513 nm) to red (705 nm). Our accomplishment will benefit in developing graphene-based multicolor display, optoelectronic integration, and even electrically driven laser in future.

## Acknowledgement

This work was supported by the National Natural Science Foundation of China (Nos.10774007, 11074006, 10874011, 50732001), the National Basic Research Program of China (Nos. 2007CB613402), and the Fundamental Research Funds for the Central Universities.